\documentclass[aps,twocolumn,superscriptaddress,nofootinbib,tighten,showpacs]{revtex4}
\usepackage{graphicx} 
\usepackage{epstopdf}
\usepackage{color}
\usepackage{multirow}
\begin{document}

\title{Mass Varying Neutrinos in Supernovae}

\author{F. Rossi-Torres\footnote{ftorres@ift.unesp.br}} 
\affiliation{Instituto de F\'isica Gleb Wataghin - UNICAMP, {13083-859}, Campinas SP, Brazil}
\affiliation{Instituto de F\'isica Te\'orica, Universidade Estadual Paulista, Rua Dr. Bento Teobaldo
             Ferraz, 271 - Bl. II, 01140-070, S\~ao Paulo, SP, Brazil}

\author{M. M. Guzzo\footnote{guzzo@ifi.unicamp.br}}
\affiliation{Instituto de F\'isica Gleb Wataghin - UNICAMP, {13083-859}, Campinas SP, Brazil}

\author{P. C. de Holanda\footnote{holanda@ifi.unicamp.br}}
\affiliation{Instituto de F\'isica Gleb Wataghin - UNICAMP, {13083-859}, Campinas SP, Brazil}

\author{O. L. G. Peres\footnote{orlando@ifi.unicamp.br}}
\affiliation{Instituto de F\'isica Gleb Wataghin - UNICAMP, {13083-859}, Campinas SP, Brazil}
\affiliation{
The Abdus Salam International Centre for Theoretical Physics, I-34100 Trieste,
Italy
}

\begin{abstract}
We study the consequences on the neutrino oscillation parameter space,
mixing angle ($\tan^2\theta$)
and vacuum mass difference ($\Delta m^2_0$),
when mass varying neutrino (MaVaN) models are assumed in a supernova
environment. We consider electronic to sterile channels $\nu_e
\rightarrow \nu_s$ and $\bar\nu_e \rightarrow \bar\nu_s$ in two-flavor
scenario. In a given model of MaVaN mechanism, we induce a
position-dependent effective mass difference, $\Delta \tilde m^2(r)$,
where $r$ is the distance from the supernova core, that changes the neutrino and
anti-neutrino flavour conversion probabilities.  
We study the constraints on the mixing
angle and vacuum mass difference coming from r-process and the SN1987A
data. Our result is the appearance of a new exclusion region 
for very small mixing angles, $\tan^2\theta=10^{-6}-10^{-2}$, and
small vacuum mass difference, $\Delta m^2_0=~1-20$~eV$^2$, due the MaVaN
mechanism.
\end{abstract}
\pacs{13.15.+g, 14.60.Pq, 14.60.St}
\maketitle

\section{Introduction}\label{intro}

Evidences from experimental data of type Ia supernovae
(SNIa)~\cite{Riess:2004nr,Astier:2005qq}, cosmic microwave background
(CMB) radiation~\cite{Spergel:2003cb} and large scale
structure (LSS)~\cite{Tegmark:2006az} point out that our universe is
in accelerated expansion. One possible explanation for this acceleration
is dark energy which contains 70\% of the total energy of our universe. 
Dark energy could be a cosmological constant ($\Lambda$), or simply, the 
non-zero vacuum energy, which is quite close to the critical cosmological energy
density, $\rho_c \approx \rho_\Lambda \approx 4 \times
10^{-47}\mbox{GeV}^4$. However one critical point of $\Lambda$ is the
necessity of a fine tuning, once that according to theoretical
expectations, $\rho_{vac}$ is $10^{50}-10^{120}$ larger than
the magnitude allowed in cosmology. In reference \cite{Dolgov:2004xu},
Dolgov emphasizes:

\begin{enumerate}

\item {\it Why vacuum energy, which must stay constant in the course
  of cosmological evolution, or dark energy, which should evolve with
  time quite differently from the normal matter, have similar
  magnitude just today, all being close to the value of the critical
  energy density?}

\item {\it If universe acceleration is induced by something which is
  different from vacuum energy, then what kind of field or object
  creates the observed cosmological behavior? Or could it be a
  modification of gravitational interactions at cosmologically large
  distances?}

\end{enumerate}

From the questions above, other candidates for dark energy have been
proposed, such as: quintessence, K-essence, tachyon field, phantom
field, dilatonic dark energy, Chaplygin gas (for excellent reviews see
\cite{reviews}.)\footnote{There are alternative scenarios to explain
  the acceleration, such as modified gravity with an
  introduction of quantum effects - higher curvature corrections - to
  the Einstein-Hilbert action.}.

Furthermore, neutrinos are powerful tools for astronomy investigation
\cite{Vissani:2009vv} and they play a crucial role in the mechanism of
supernova explosion, according to the standard
scenario~\cite{Bethe:1984ux}.  Based on the very interesting fact that
the scale of the neutrino squared mass difference
($(0.01~\mbox{eV})^2$) is similar to the dark energy scale, the
possibility of neutrino coupling with a scalar field naturally
arises. Hung \cite{hung} developed the idea that sterile neutrinos
could have a relation with the accelerating universe.  He proposed
that the sterile neutrino obtains its mass through a Yukawa coupling
with a singlet scalar field whose effective potential is of a
``slow-rolling'' type in which the vacuum energy is given
approximately by $\rho_{vac}$ and the effective mass of the sterile
neutrino is proportional to the expectation value of the scalar
field. When this scalar field evolves, the sterile neutrino mass
changes. Fardon~{\it et al.} describe that the total energy of the
cosmological fluid can vary slowly as the neutrino density
decreases~\cite{Fardon:2003eh}, in a similar way of Hung's model.
Consequently, the coupling with a scalar field causes a variation of
the neutrino mass (dynamical field) proportional to the local neutrino
density and perhaps to the baryonic matter~\cite{Kaplan:2004dq}. This
picture has been called Mass Varying Neutrino (MaVaN) models 
. Afshordi~{\it et al.}~\cite{Afshordi:2005ym} and Bjalde~{\it et
al.}~\cite{Bjaelde:2007ki} showed that the proposed MaVaN models, in
the non-relativistic regime, could be threatened by the strong growth of
hydrodynamic perturbations associated with a negative adiabatic squared sound
speed. Many articles have treated neutrino oscillation via the
MaVaN mechanism~\cite{Barger:2005mn,Cirelli:2005sg,GonzalezGarcia:2005xu,Gu:2005eq,Gu:2005pq,Schwetz:2005fy,Abe:2008zza,Zurek:2004vd,holanda}.
We observe, nevertheless, that these articles focus on solar and
atmospheric neutrinos, as well as observations from accelerators and
reactors. However, the assumption of MaVaN mechanism in the neutrino
evolution in supernovae environments has not been fully
explored yet. In fact, the only article that connects MaVaN and supernova
is Li~{\it et al.}~\cite{Li:2005zd} which use data from a type Ia
supernova to limit the interactions between neutrinos and scalar dark
energy.

In this paper, therefore, we investigate the consequences of MaVaN
mechanisms on the neutrino propagation in a supernova environment. We
take into consideration that recent results from
WMAP-7~\cite{Komatsu:2010fb} indicate that the number of species of
relativistic neutrinos is equal to $4.34 \pm 0.87$ which, in contrast
with the standard model three active neutrinos, suggest that there may
be additional species of neutrinos, for example, a sterile one
$\nu_s$, possibly an SU(2) singlet. These sterile neutrinos can
present masses ranging from a few eV to values larger than the
electroweak scale~\cite{Kusenko:2007wv}. Therefore, a natural and
potentially interesting investigation can arise from the analyses of
(anti)neutrino oscillations $\nu_e \rightarrow \nu_s$ and $\bar\nu_e
\rightarrow \bar\nu_s$ in supernovae under the assumption that MaVaN
models are implemented. We evaluate the consequences of such (anti)neutrino
oscillations on the signal of $\bar\nu_e$ and $\nu_e$ in terrestrial
detectors, as well as on supernova heavy-element nucleosynthesis (r-process).

The crucial point of our analysis is the fact that MaVaN model can
directly affect the relevant parameter for neutrino oscillations,
namely, the squared mass difference, $\Delta m^2_0$. We propose a
phenomenological MaVaN model that is more convenient to test using
neutrinos from supernova. Such parameterization allows to fit the
amplitude of the $\Delta m^2_0$ variation, the position where the
variation begins and also if the masses will increase or decrease
along the neutrino
trajectory. We use this parameterization to evaluate the constraints
on the oscillation parameter space, $\tan^2\theta \times \Delta m^2_0$
(where $\theta$ and $\Delta m^2_0$ are the mixing angle and the squared
mass difference in vacuum, respectively), coming from the r-process
nucleosynthesis condition, $Y_e<0.5$~\cite{nuas,Nunokawa:1997ct}, and
by the limit for the average survival probability coming from SN1987A
data, $\langle P \rangle <0.5$. Note that problems and criticisms to
these MaVaN models were pointed by Peccei in Ref.~\cite{Peccei:2004sz}.

The conclusion of our analysis is the following. With our MaVaN parameterization, 
significant modifications happen in the probability iso-curves in (anti)neutrino 
parameter space, leading to new regions of exclusion. This happens 
particularly when the evolution of mass-squared difference inside the star is 
suficiently slow to avoid non-adiabatic conversions. Besides, we find that 
the r-process nucleosynthesis is not affected by the MaVaN assumption.

This article is organized as follows: section \ref{osc} presents the
neutrino oscillation mechanism and the assumptions and approximations
used in our work. Section \ref{mavan} presents our proposition of a
new parameterization of MaVaN model. Section \ref{results} contains our
results and related discussion. Finally section \ref{conclusion}
summarizes our conclusions.

\section{Neutrino Oscillation and Supernova Neutrino Spectrum}\label{osc}

Supernova flavor spectra are not known with precision. In fact, several
degeneracies are presented in neutrino emission parameters, as it is shown in
\cite{minakata}. Therefore, we will focus on an unpretentious approach:
how the pattern of flavor conversion is affected by the inclusion of MaVaN
in a supernova environment. In this framework, we assume a two-neutrino
oscillation phenomenon and that some typical emission spectrum can be
representative of a wide range of possibilities. Although this is a very
naive approach, it is sufficient to appreciate the general features of the
important modifications in the neutrino oscillation pattern introduced by
MaVaN in a supernova environment.

Some comments about the assumptions for the neutrino oscillations we
are interested in this article are in order. From the neutrino
phenomenology we know the existence of two small mass scales, $\Delta
m^2_{atm}$ and $\Delta
m^2_{sun}$ and two large mixing angles, $\theta_{21}$ and
$\theta_{23}$, and one small angle, if not vanishing,
$\theta_{13}$. We are interested in oscillations between active
electronic into sterile neutrinos as well as oscillations between
active electronic into sterile anti-neutrinos in a supernova
environment. For this purpose we will include a new scale, {\bf $\Delta
m^2_0$}. This new scale needs to be much larger than the small scales for
active neutrinos - $\Delta m^2_0>>\Delta m^2_{atm,sun}$ - in order to be
compatible with constraints for sterile neutrinos.\footnote{For an analysis of
sterile neutrino oscillations in cosmological, astrophysical and
terrestrial media with various mixing and a wider range of $\Delta
m^2_0$ see~\cite{Cirelli:2004cz}.}

Such oscillations must be treated considering that both neutrino and
anti-neutrino conversion can be resonantly enhanced in matter, the
so-called Mikheyev-Smirnov-Wolfenstein (MSW) phenomenon~\cite{msw}.
Nevertheless, for a convenient choice of the squared mass difference
in vacuum, we are allowed to neglect the conversion to sterile
neutrinos in the inner core, for all values of the mixing angles we
considered~\cite{Kainulainen:1990bn}.  We will assume therefore that
$1 \mbox{eV}^2<\Delta m^2_0<10^4 \mbox{eV}^2$, always positive.
Consequently, neutrinos carrying tens of MeV of energy will not suffer
any resonance, since the core matter density is approximately
10$^{14}$~g/cm$^3$.  Then the electronic potential is so high that no
resonance will occur for any mixing angle inside the core.
Furthermore, one neglects $\nu-\nu$ forward-scattering contribution to
the weak potentials in first approximation after neutrinos have
escaped from the inner core of the star.  Recently, issues of self
neutrino interactions were discussed considering three families by
Dasgupta~\cite{amol} and Friedland~\cite{friedland}. An analysis of
self neutrino interactions considering two families was done
in~\cite{duan,dasgupta}.

The equation which governs the flavor neutrino eigenstates evolution
along their trajectory inside the supernova can be written as:
\begin{equation} i\frac{\partial}{\partial r} \left[
\begin{array}{c} \Psi_e(r)\\ \Psi_s(r)
\end{array} \right]=\left[
\begin{array}{cc} \phi_e(r) & \sqrt{\sigma}\\ \sqrt{\sigma} &
-\phi_e(r)
\end{array} \right]\left[
\begin{array}{c} \Psi_e(r)\\ \Psi_s(r)
\end{array} \right],
\label{eq7}
\end{equation} where
\begin{equation} \phi_e(r)=\frac{1}{4E}(\pm 2V(r)E-\Delta m^2_0\cos
2\theta),
\label{eq8}
\end{equation} and
\begin{equation} \sqrt{\sigma}=\frac{\Delta m^2_0}{4E}\sin 2\theta.
\label{eq9}
\end{equation} 
with $\Delta m^2_0>0.$
Under the assumptions described above, the matter
potential can be written as follows
\begin{equation} V(r) = \sqrt{2} G_F
\left[N_{e^-}(r)-N_{e^+}(r)-\frac{N_n(r)}{2} \right].
\label{eq10}
\end{equation} In Eq. (\ref{eq8}), the signal $+$ is for neutrino,
whereas the $-$ signal is for anti-neutrinos. $\Delta m^2_0$ is the
squared mass difference between two neutrinos mass eigenstates in
vacuum; $G_F$ is the Fermi coupling constant; $\theta$ is the vacuum
mixing angle and $N_{e^-}(r)$, $N_{e^+}(r)$ and $N_n(r)$ are,
respectively, the number densities of electrons, positrons and
neutrons. $N_p=N_{e^-}-N_{e^+}$, which is the number density of
protons, does not appear in Eq.~(\ref{eq10}), because electric
charge neutrality of the medium was assumed.  Note that neutrinos (anti-neutrinos)
undergo in a resonance when
\begin{equation} V(r)=\pm\frac{\Delta m^2_0}{2 E} \cos 2\theta.
\label{eq12}
\end{equation}

The range for $\theta$ parameter is $0\le\theta\le \pi/2$. Our results 
can also be extended for the neutrino case oscillation, where the results
in the region of $0\le\theta\le \pi/4$ for anti-neutrinos will be the
same for the neutrinos in the region of $\pi/4\le\theta\le \pi/2$ and
vice-versa. 

The matter potential can be rewritten as
\begin{equation} V(r)=\frac{3G_F
\rho(r)}{\sqrt{2}m_N}\left(Y_e-\frac{1}{3} \right),
\label{eq14}
\end{equation} where $Y_e$ is the electronic fraction
\begin{equation} Y_e(r)=\frac{N_{e^-}(r)-N_{e^+}(r)}{N_p(r)+N_n(r)},
\label{eq13}
\end{equation} $m_N$ is the nucleon mass and $\rho$(r) is the matter
density profile.

We will consider that the spectrum of neutrino emission can be written
as~\cite{Giuntibook}:
\begin{equation} \frac{dN}{dE}=\frac{L}{F(\eta)
T^4}\frac{E^2}{e^{E/T-\eta}+1}~~~,
\label{espec}
\end{equation} where $\eta$ is the pinching factor and
$F(\eta)~=~\int_0^\infty dx x^3/(e^{x-\eta}+1)$.  For this spectrum
$\langle E \rangle/T \approx 3.1514 +
0.1250~\eta+0.0429~\eta^2+\mbox{O}(\eta^3)$. Typical values of $\eta$
are $\eta_{\nu_e} \sim 2$, $\eta_{\bar\nu_e} \sim 3$ and $\eta_x \sim
1$~\cite{Keil:2002in}. The average energy of neutrino species is
$\langle E_{\nu_e} \rangle \approx 11 $~MeV, $\langle E_{\bar\nu_e}
\rangle \approx 16 $~MeV, $\langle E_{\nu_x} \rangle \approx 25 $~MeV
($x~=~\mu,\tau$). $L$ is the neutrino luminosity, which is
approximately $10^{51}$~erg s$^{-1}$ and it can be considered equal
for all neutrino flavors. Other choice of parameter
spectra in Eq.~(\ref{espec}), using codes such as those ones presented in 
\cite{totani,garching}, can be extracted.
We use in this work the electronic
potential profile for a supernova with a post bounce time
evolution equal to $t_{pb}~=~2$~s, where the size of the
neutrinosphere is approximately equal to the core size and the
supernova is in the stage where heavy nuclei start to form.

In the MSW effect, the computation of the neutrino probabilities can
have two regimes, the adiabatic regime which depends only on mixings
of neutrinos in the initial and final point (and then
can be calculated analytically) or the non-adiabatic regime, which depends
on the path crossed by the neutrinos and can be found by solving
numerically the Eq.~(\ref{eq7}). In standard scenario, all adiabatic effects
for small mixings happen for $\Delta m^2_0 \gtrsim 100$~eV$^2$. For the 
results shown here, we will use the averaged probability over the neutrino 
spectrum described in Eq.~(\ref{espec}).

\section{Mass Varying Neutrino model}\label{mavan}
 
We adopt a phenomenological approach in modeling the MaVaN
mechanism\footnote{In~\protect{\cite{Franca:2009xp}} it was proposed a
phenomenological construction of a MaVaN model in the cosmological
context.}.  We will implement the MaVaN mechanism by changing the
mass-squared difference in vacuum ($\Delta m^2_0$) to a mass-squared
difference ($\Delta \tilde m^2$) that depends on the neutrino density
($n_\nu$) of the environment. In principle,  we can also change the
mixing angle, but we prefer to avoid this for simplicity.  Also, the
squared mass difference in MaVaN framework is the same for neutrinos
and anti-neutrinos, $(\Delta \tilde{m}^2)_{\nu}~=~(\Delta
\tilde{m}^2)_{\bar{\nu}}$.

We choose a parameterization for squared mass difference ($\Delta
\tilde m^2$) in the MaVaN scenario such that we have, {\it at same
time}, MaVaN induced effects and Mikheyev-Smirnov-Wolfenstein (MSW)
effects~\cite{msw}.
For this we build the
following function
\begin{equation} 
\Delta \tilde{m}^2(r) ~=~ \Delta m^2_0 -
\frac{\delta}{1+(n_\nu (r)/n_\nu^0)^{-\eta}}~~,
\label{eq:paramet}
\end{equation} where the neutrino density, $n_\nu (r)$, 
assuming that the neutrinos
are equally produced in the neutrinosphere 
with radius $R_\nu \sim 10$~km, is given by
\begin{equation} n_\nu (r) ~=~ \frac{L}{\langle E_\nu \rangle}
\frac{1}{8\pi c R_\nu r} \ln \left(\frac{r+R_\nu}{r-R_\nu}\right),
\label{eq:nuden}
\end{equation} where $r$ is the
radial distance measured from $R_\nu$ and $c$ the speed of light.  For
larger radius $r \gg R_\nu$, the neutrino density is written as
$n_{\nu}~\sim r^{-2}$.  The parameters $n_\nu^0$ and $\eta$ are constants
to be chosen. For $n_\nu \ll n_\nu^0$ ($n_\nu \gg n_\nu^0$) the
value of $\Delta\tilde{m}^2$ tends to the asymptotic value of $\Delta
m^2$ ($\Delta m^2_0-\delta$). The choice of $n_\nu^0$ and $\eta$ are
based in some considerations about the relative weight of MaVaN and
the MSW mechanism.
\begin{figure}[ht]
\begin{center}
\includegraphics[width=0.97\linewidth]{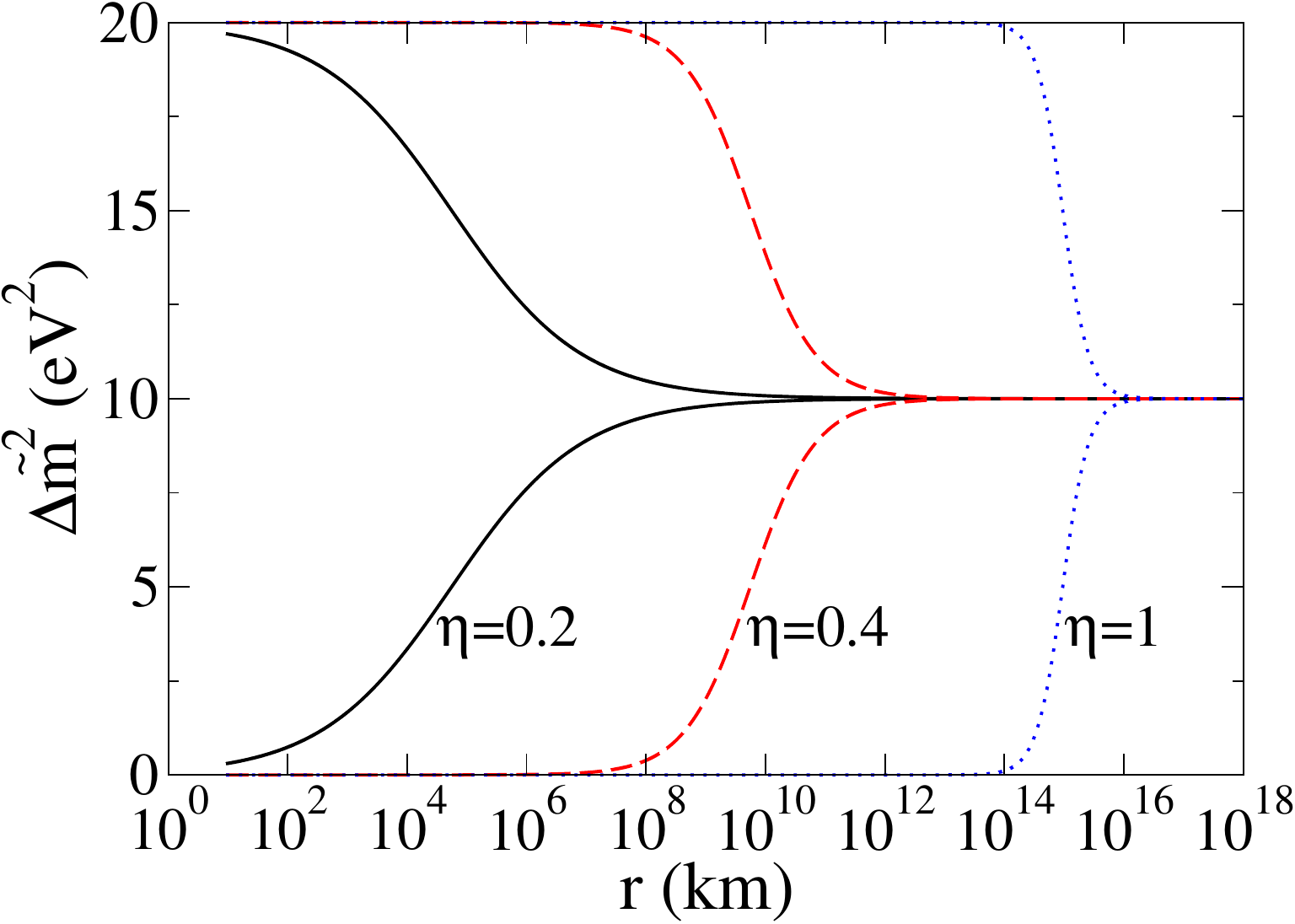}
\caption{\small \it (Colors in online edition) The evolution of
$\Delta \tilde{m}^2$ from Eq.(\ref{eq:paramet}) is shown for $\Delta
m^2_0~=~10$ eV$^2$, $\delta~=~\pm 10$ eV$^2$ and $n_\nu^0~=~10^{25}$,
$10^{15}$ and $34500$~cm$^{-3}$ for the solid, dashed and dotted
lines, respectively. The values for the parameter $\eta$ are shown in
the plot. Upper curves for negative $\delta$ and lower curves are for
positive $\delta$.  
}
\label{fig:massparamet}
\end{center}
\end{figure}

In Fig.~\ref{fig:massparamet} we plot the effective mass difference,
$\Delta \tilde{m}^2$, for some values of $n_\nu^0$ and $\eta$. For the
choice of the neutrino luminosity $L\sim 10^{51}$~erg s$^{-1}$ and
neutrino average energy $\langle E_{\nu_x} \rangle
\approx 16 $~MeV, we have that for almost any value above
$n_\nu^0~>~10^{28}$~cm$^{-3}$ and $\eta<1$ the effective mass difference
$\Delta\tilde{m}^2$ reaches the vacuum mass difference 
$\Delta m^2_0$, {\it inside the supernova}. For
comparison, the model presented in~\cite{Cirelli:2005sg} is very well
reproduced by choosing $\delta~=~10$ eV$^2$, $\eta~=~1$ and
$n_\nu^0~=~34500$~cm$^{-3}$ but for these parameters we have no
overlap for MaVaN and MSW effects.

The MaVaN mechanism with MSW effects will be effectively given by 
changing $\Delta m^2_0\to \Delta\tilde{m}^2$ in
Eqs. (\ref{eq7})-(\ref{eq12}). The interplay between the 
MaVaN effect with MSW effect is very effective when the resonance condition
happens for the effective mass difference
\begin{equation} V(r)=\pm\frac{\Delta \tilde{m}^2(r)}{2 E} \cos 2\theta.
\label{eq12mavan}
\end{equation}
with the +(-) sign attributed to neutrinos (anti-neutrinos). In the MaVaN mechanism, due the
$n_{\nu}~\sim r^{-2}$ asymptotic behavior, we can have the 
$\Delta \tilde{m}^2(r)$ changing the sign inside the supernova, 
from positive to negative values, including the point where
$\Delta \tilde{m}^2(r)\rightarrow 0$ as shown in the bottom part of 
Fig.~\ref{fig:massparamet}. In this region non-adiabatic matter effects 
can happen.  

\begin{figure}[h]
\begin{center}
\includegraphics[width=0.97\linewidth]{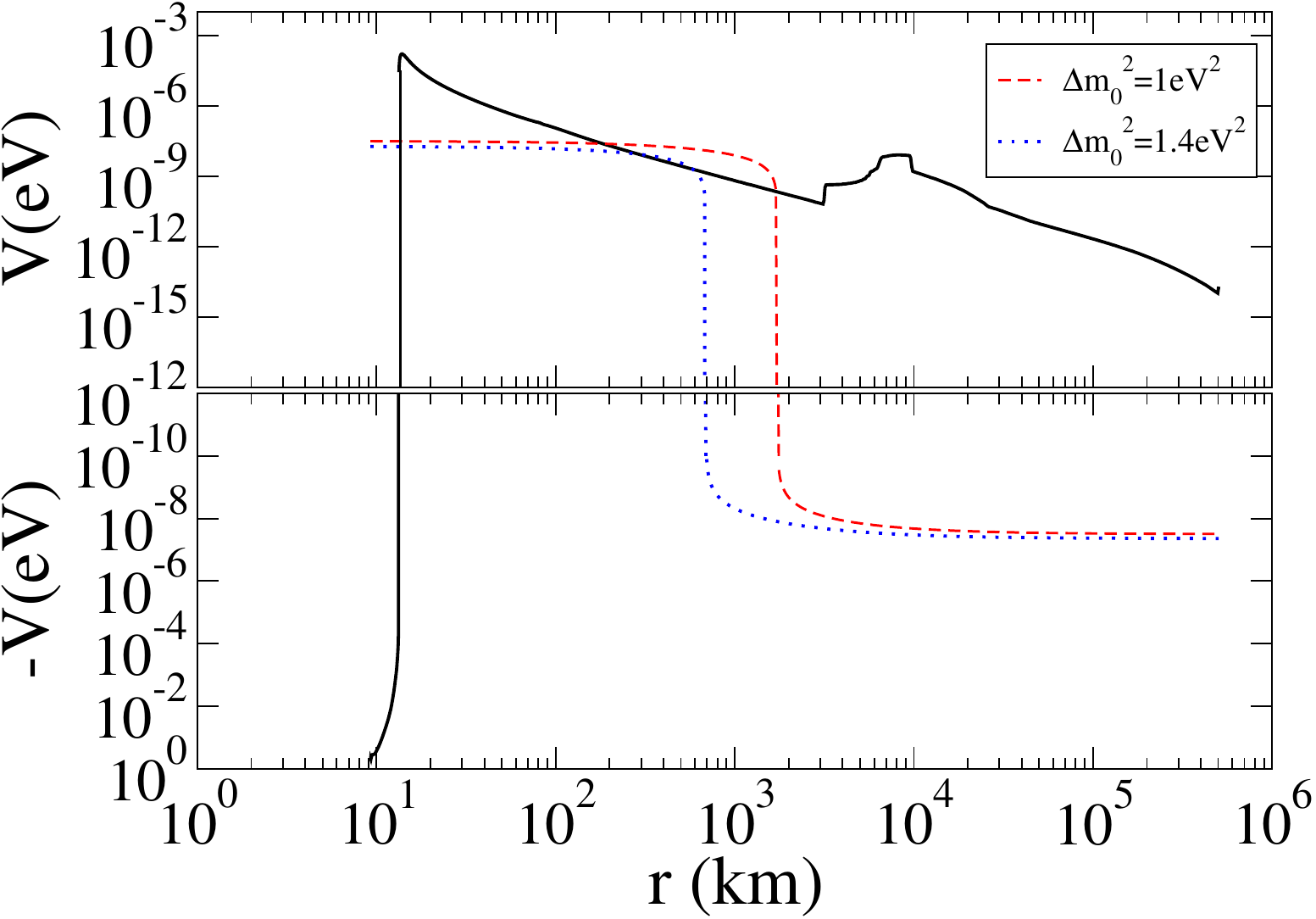}
\caption{\small \it (Colors in online edition) The solid curve
represents the electronic potential ($V_e$), calculated from
Eq.~(\ref{eq14}). Dotted (dashed) curve is the right-side of 
Eq.~(\ref{eq12mavan}) for anti-neutrinos, for $\Delta m^2_0=1.4~(1)$~eV$^2$, 
with $\delta~=~2$~eV$^2$ and
($\eta$,$n^0_\nu$)~=~($0.5$,$1\times10^{28}$~cm$^{-3}$) for
$\tan^2\theta ~=~ 2.5 \times 10^{-5}$ and $\langle E_\nu
\rangle~=~15$~MeV.}
\label{fig:ress}
\end{center}
\end{figure}

Fig.~\ref{fig:ress} shows
the electronic potential ($V_e$, solid line) and 
the right side of Eq.~(\ref{eq12mavan}) for anti-neutrinos, 
with the minus sign, for 
$\Delta m^2_0=1$~eV$^2$ (dashed line) and
$\Delta m^ 2=1.4$~eV$^2$ (dotted line), for a value of
$\tan^2\theta=2.5 \times 10^{-5}$ and $\delta=2$ eV$^2$ with an
average neutrino energy $\langle E_\nu \rangle=15$~MeV. We have
chosen the $\delta$ parameter inside the range 
$\delta=1-20$~eV$^2$, so that the effect of new physics does not disturb
the adiabatic effects that happens for $\Delta m^2_0 \gtrsim 100$~eV$^2$, as 
discussed before. In this way the new MaVaN effects can be analysed more 
independently, avoiding a disturbance in the standard adiabatic effects in the
inner region of supernova for large values of $\Delta m_0^2$, around a radial 
distance of 10~km. In the MaVaN framework, from Fig~\ref{fig:ress}, it is possible to
see the appearance of new resonance points. 
Now we have fixed the range for the parameters
$\delta$ and $n_\nu^0$ and we can begin to study the MaVaN phenomenology.

\section{Results and Discussion}\label{results}

It is possible to constrain the oscillation parameters, such as the
mixing angle and the squared mass difference, from the SN1987A data
\cite{k,i,b} in a number of ways. One of them comes from the spectrum
of the observed events, probably dominated by $\bar\nu_e$. From the
observed energy spectrum of SN1987A, the analysis from Kamiokande
indicate that the temperature of $\bar\nu_e$ is lower than the
expectation~\cite{murayama}. The second constraint comes from the
expanding envelope driven by thermal neutrino wind of the supernova
which is a possible site of heavy nuclei formation beyond iron
(r-process nucleosynthesis). Since the $\nu_e$ and $\bar\nu_e$
conversion into a sterile state (or even into an active state) happens
in different rates, the fraction of neutrons to protons, determined by
$\bar\nu_e + p \rightarrow e^+ + n$ and $\nu_e + n \rightarrow e^- +
p$, is modified by the oscillation mechanism in a radial distance from
the center of the star of about a hundred kilometers. Related to
this constraint, a very simple bound is $Y_e < 0.5$, which can be
extracted from Eq.~(\ref{eq13}) simply considering a very neutron-rich
environment~\cite{nuas,Nunokawa:1997ct}. In~\cite{Nunokawa:1997ct} it
is shown that $Y_e$ has a dependence on the neutrino oscillations
probabilities given by
\begin{equation} Y_e \sim \frac{1}{1+P_{\bar\nu}\langle E_{\bar\nu_e}
\rangle /P_\nu\langle E_{\nu_e} \rangle},
\end{equation} 
where $P_{\bar\nu}$ and $P_\nu$ are, respectively, the survival
probabilities of anti-neutrinos and neutrinos. From this expression,
we clearly see that it is possible to constrain the oscillation
parameters from $Y_e < 0.5$. In \cite{murayama}, it is also discussed
another constraint based on the fact that the first or the second
event of Kamiokande could be related with $\nu_e$. Recent analysis of SN1987A 
data can constrain the $\bar\nu_e$ supernova emission model~\cite{pagliaroli}.

We divide this section into two parts. The first one concerns the
analysis of the oscillation of active electronic (anti)neutrinos to
sterile (anti)neutrinos. In the second section we discuss the possible
limits imposed by r-process nucleosynthesis in the oscillation
parameter space. In these subsections we will compare the case of
MaVaN scenario, presented in section \ref{mavan}, with the case of
standard oscillations.

\subsection{(Anti)neutrinos}\label{neut}

Here we calculate the survival probability $P_{\bar\nu}$
for the $\bar\nu_e \rightarrow \bar\nu_s$ channel of oscillation.
Based on the data from SN1987A and
neutrino oscillation only in two families, several articles, 
such as~\cite{Nunokawa:1997ct}, discuss that the survival probability cannot
be less than $0.5$, for a $\bar\nu_e \rightarrow \bar\nu_s$
conversion.  In this way, it is possible to find regions of exclusion
in the oscillation parameter space ($\Delta m^2_0,\tan^2\theta$). 

We will find probabilities in the context of the MaVaN
parameterization, given by Eq.~(\ref{eq:paramet}), where 
we have chosen the sets for ($\eta$, $n_\nu^0$) as given in 
Table~\ref{tab}.
\begin{table}[t]
\begin{center}
\begin{tabular}{ccccc} \hline\hline case & $\eta$ & $n_\nu^0$ & $\delta$ & Figure \\ \hline
\multirow{2}{*}{case~1}  & $0.5$ & $1.0\times 10^{28}$ &$+20$~eV$^2$ & Figure~\ref{fig3}\\ 
& $0.5$ & $1.0\times 10^{28}$ &$-20$~eV$^2$ & Figure~\ref{fig5}\\ \hline 
case~2 & $0.8$ &
$1.5\times 10^{30}$ &$+20$~eV$^2$ & Figure~\ref{fig4}\\ \\ \hline 
\end{tabular}
\caption{\small \it Three sets ($\eta$,$n_\nu^0,\delta$) of our MaVaN
model.}\label{tab}
\end{center}
\end{table}

\begin{figure}[h]
\begin{center}
\includegraphics[width=0.97\linewidth]{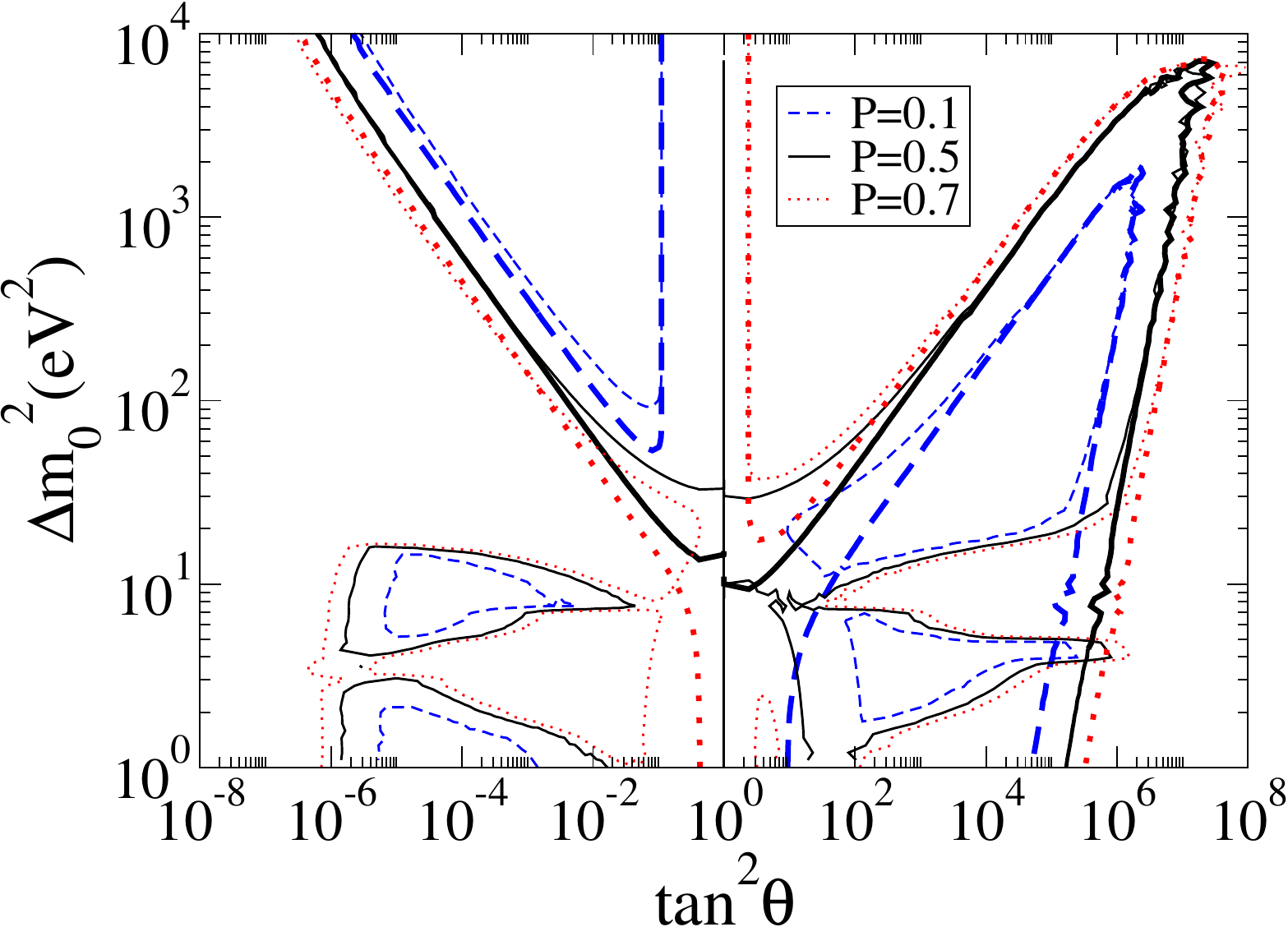}
\caption{\small \it (Colors in online edition) Isocurves of survival
probability ($P$). The dashed represents $P~=~0.1$, the solid one
$P~=~0.5$ and the dotted curve $P~=~0.7$. The curve evolution set
($\eta$, $n_\nu^0$) ~=~ ($0.5,1\times10^{28}$~cm$^{-3}$) (case 1)
for $\delta ~=~20~eV^2$. In this figure, the thicker curves
represent the case without MaVaN ($\delta~=~0$).}
\label{fig3}
\end{center}
\end{figure}

In Fig.~\ref{fig3} we show the isocurves of average probability in
the oscillation parameter space ($\Delta m^2_0,\tan^2\theta$) with MaVaN
(in this figure we use the first line of the Case 1 in
Table~\ref{tab}: ($\eta$, $n_\nu^0$) ~=~ ($0.5,1\times10^{28}$~cm$^{-3}$) and 
$\delta=+20$~eV$^2$) and without MaVaN. The thicker curves in
Fig.~\ref{fig3} are for the situation without MaVaN and thinner 
ones represent the situation with MaVaN. In the
standard scenario without MaVaN, for lower masses $\Delta m^2_0\ll
10$~eV$^2$ and $\tan^2\theta~\ll 10^{-1}$, the survival probability is
greater than 0.7 and then in agreement with SN1987A data. When we
include MaVaN, with the parameters in the Case 1 of Table~\ref{tab},
we have the presence of new area in the region of
$\tan^2\theta~\approx~10^{-6}-10^6$ and $\Delta m^2_0~\approx~
1-10$~eV$^2$ with survival probabilities smaller than 0.5 in direct
contradiction with SN1987A data. Then in this case we can exclude
oscillation parameter regions that cannot be tested in the standard
scenario. The isocontours are different depending on whether
the resonance condition is being fulfilled or not.

To explain the appearance of these new regions we should
resort to Fig.~\ref{fig:massparamet}. The effective mass difference 
in MaVaN mechanism, $\Delta\tilde{m}^2(r)$, can change sign 
inside the supernova, leading to new resonances that were not present before.
In standard oscillation scenario without MaVaN, the right
side of Eq.~(\ref{eq12}) is always negative and the resonance
condition is never fulfilled. But in the MaVaN
mechanism, the effective mass difference $\Delta
\tilde{m}^2$ change sign inside the supernova, as shown in
Fig.~\ref{fig:ress}, and now it is possible to fulfill the condition of
Eq.~(\ref{eq12mavan}).
\begin{figure}[h]
\begin{center}
\includegraphics[width=0.97\linewidth]{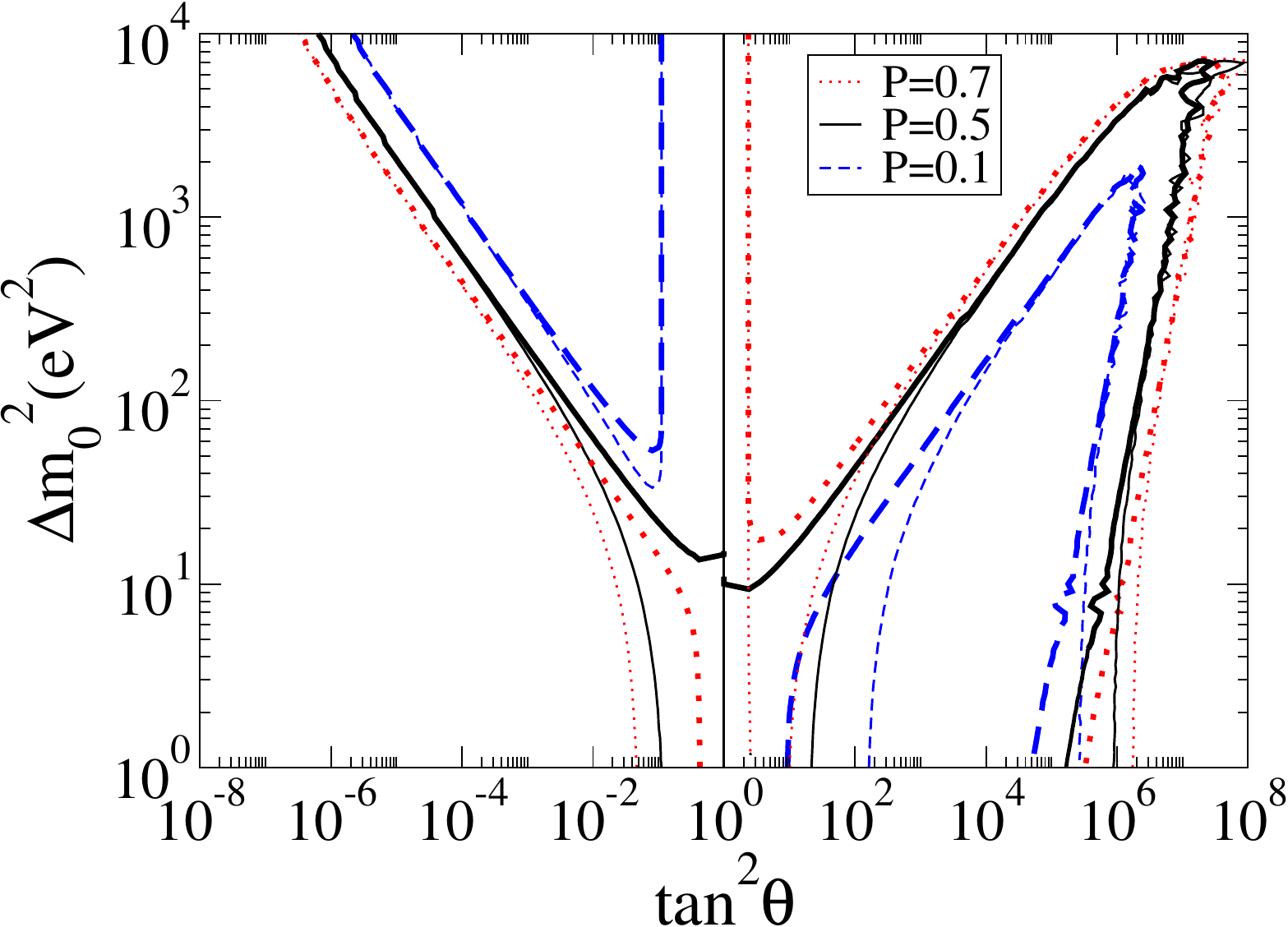}
\caption{\small \it (Colors in online edition) Isocurves of survival
probability ($P$). The dashed represents $P~=~0.1$, the solid one
$P~=~0.5$ and the dotted curve $P~=~0.7$. The curve evolution set
($\eta$, $n_\nu^0$) ~=~ ($0.5,1\times10^{28}$~cm$^{-3}$) (case 1)
for $\delta ~=~-20~eV^2$. In this figure, the thicker curves
represent the case without MaVaN ($\delta~=~0$).}
\label{fig5}
\end{center}
\end{figure}

To test the robustness of our analysis, we choose a negative value for
$\delta$ parameter ($\delta=-20$~eV$^2$) as described in the second
line of the Case 1 in Table~\ref{tab}: ($\eta$,
$n_\nu^0$)~=~($0.5,1\times10^{28}$~cm$^{-3}$).  The results are shown
in Fig.~\ref{fig5} by the isocontours with MaVaN (thinner curves) and
without MaVaN scenario (thicker curves). Although there are changes of
the isocontour curves of survival average probability with MaVaN
compared with the usual scenario without MaVaN, they are smaller than
the case of positive $\delta$. Then we can conclude that only a
positive $\delta$ will change significantly the anti-neutrino
survival probability and then could be ruled out by SN1987A data. For
smaller values of $\delta$, the MaVaN mechanism is even less important
and there is a very small modification in the isocontours of survival
average probability if we compare with the standard scenario.
  
\begin{figure}[h]
\includegraphics[width=0.97\linewidth]{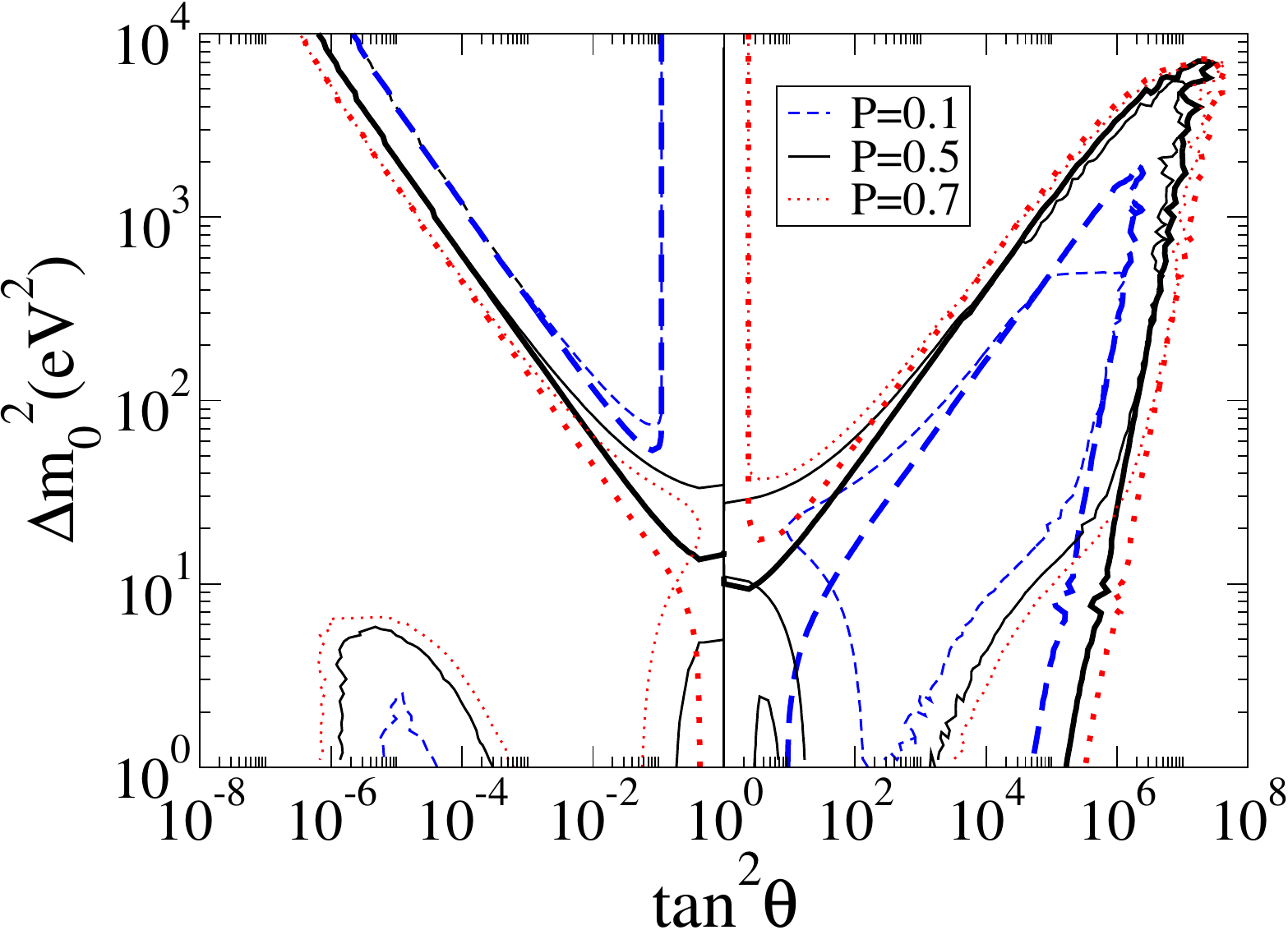}
\caption{\small \it (Colors in online edition) Isocurves of survival
probability ($P$). The dashed represents $P~=~0.1$, the solid one
$P~=~0.5$ and the dotted curve $P~=~0.7$. The curve evolution set
($\eta$, $n_\nu^0$) ~=~ ($0.8,1.5\times10^{30}$~cm$^{-3}$) (case 2)
for $\delta~=~20~eV^2$. In this figure, the thicker curves
represent the case without MaVaN ($\delta~=~0$).}
\label{fig4}
\end{figure}

Another test is obtained by changing the parameter $n_\nu^0$ using the
Case 2 in Table~\ref{tab} as shown in Fig.~\ref{fig4}, where
($\eta$,$n_\nu^0$)=($0.8,1.5\times10^{30}$~cm$^{-3}$) and
$\delta=+20$~eV$^2$.  For this choice with higher values of $n_\nu^0$
or $\eta$, the values of the isocontour curves of the average
probability in the bottom part of Fig.~\ref{fig4} shrink to smaller
values of vacuum mass differences and mixing angles if
compared to the situation shown in Fig.~\ref{fig3}. For even higher
values of ($\eta$,$n_\nu^0$), new isocontour regions are even smaller
(negligible) and closer to the $\delta=0$ solution, because the
condition of Eq.~(\ref{eq12mavan}) is reached only in the inner parts
of supernova, where non-adiabatic effects are dominant and then the
survival probability is near the maximum.

From these tests, we conclude that in the MaVaN mechanism is possible
to change the isocontours of survival probability as shown in
Figs.~\ref{fig3} and \ref{fig4} for the optimal values of MaVaN
parameters as described in Case 1 of Table~\ref{tab}. 
If we remove
some of our assumptions: the coexistence of MaVaN and MSW effects and
the adiabaticity of inner resonance, then we can have a different
range for MaVaN parameters, hence the allowed region for the
oscillation also will change.

\subsection{r-process}\label{rprocess}

\hspace{0.5cm}The production of heavy nuclei in supernova is still an
open problem. The region between the protoneutron star and the
escaping shock wave a few seconds after the bounce may be a good site
for this process, with high entropy and an excess number of
neutrons. From $\beta$ reactions, $\bar\nu_e p \rightarrow n e^+$ and
$\nu_e n \rightarrow p e^-$, a modification in the $n/p$ fraction
could happen if one considers neutrino oscillation \cite{nuas,qian}.
We will analyse what happens with this condition in the context of our
MaVaN parameterization and for $\bar\nu_e \rightarrow \bar\nu_s$ and
$\nu_e \rightarrow \nu_s$. We limit our analysis of neutrino flavor
conversion to about $r<50$~km, i. e., a region where there is not any
shock wave and has a higher influence on anti-neutrinos. This is
approximately the region where the r-process nucleosynthesis happens.
For a review of this subject see \cite{rprocessreview}.

From Fig.~\ref{fig:ress} one can observe that resonances can occur inside the
nucleosynthesis region ($r\approx 50$~km). Nevertheless such resonances are
extremely non-adiabatic when small values of $\Delta m_0^2$ are taken into
consideration. Therefore, they will not affect the relevant nuclesynthesis.
Furthermore, if one considers large values of $\Delta m_0^2$, no resonance
happens inside the relevant nucleosynthesis region. In conclusion, our
parameter choice implies that one does not expect any significant change in
nucleosynthesis processes.

The evolution of $\Delta \tilde m^2$ represented by the dotted line in
Fig.~\ref{fig:massparamet} and its implication in the r-process
nucleosynthesis and oscillation probabilities were discussed in
\cite{RossiTorres:2010zz}.

\section{Conclusions and Outlook}\label{conclusion}

In this paper we proposed a new phenomenological
parameterization for the variation of the relevant neutrino
oscillation parameters generally presented in MaVaN models, characterized
by the parameters ($\eta$,$n_\nu^0$) and $\delta$.
We analysed the neutrino and anti-neutrino survival probabilities  
for the channels $\bar \nu_e \rightarrow \bar \nu_s$ and 
$\nu_e \rightarrow \nu_s$ in the two flavor context with and without MaVaN.
We studied the constraints on the mixing
angle and vacuum mass difference coming from r-process and the SN1987A
data.

Assuming that MaVaN effects and MSW effects are equally 
important and the conditions about adiabaticity of resonances are 
mantained, we imposed
a range for  MaVaN parameters ($\eta$,$n_\nu^0$) that satisfied 
$n_\nu^0~>~10^{28}$ cm$^{-3}$
and $\eta<1$. In this range we found that the 
anti-neutrino survival probability behavior change, allowing  
smaller values of survival probability - P$\sim 0.5$ - for
very small mixing angles, ~$10^{-6}-10^{-2}$ and
small vacuum mass difference, $\Delta m^2_0=~1-20$~eV$^2$ 
that are in contradiction with SN1987A data.
Then, if the MaVaN mechanism is effective with the MaVaN parameters 
in the range that we considered, 
we can rule out regions of parameter space that were allowed in standard 
scenario, without MaVaN's effects.

We also studied the r-process nucleosynthesis condition $Y_e<0.5$
which is always fulfilled, independent of the oscillation
parameters. Then this procedure failed to produce any constraint.

For the next galactic supernova we expect a much larger amount of
events coming from $\bar\nu_e$ and $\nu_e$ elastic scattering. Then we
consider that a good understanding of what happens to electronic
neutrinos and anti-neutrinos inside the supernova, in the context of
(anti)neutrino oscillations, is crucial to understand the signal that
will be detected. Any discrepancy between experimental results and
theoretical predictions will point to new physics. In special we
presented that MaVaN mechanism can distort the neutrino and anti-neutrino 
probabilities and be a source of this discrepancy.
Needless to say that, for anti-neutrinos, the sample data will be
larger and with higher statistics the probability to find new
phenomena will increase.
  
\begin{acknowledgments} 
The authors are thankful for the ICTP, FAPESP, CNPq and CAPES and
Fulbright commission for financial support. We are also thankful for the
useful comments and suggestions made by C. A. Moura. We are very
grateful for the supernova profile supplied by the
prof. H.~T-Janka. F. Rossi-Torres is also thankful for the hospitality
of Laboratori Nazionali del Gran Sasso (LNGS) were this work was
partially developed. O. L. G. Peres thanks the C.N. Yang Institute at 
Stony Brook University 
for the hospitality were this work was partially developed.

\end{acknowledgments}


\begin{thebibliography}{9}

\bibitem{Riess:2004nr} A.~G.~Riess {\it et al.}  [Supernova Search
Team Collaboration], 
 Astrophys.\ J.\ {\bf 607},
  665 (2004)
 [arXiv:astro-ph/0402512].  

\bibitem{Astier:2005qq} P.~Astier {\it et al.}  [The SNLS
  Collaboration],
  Astron.\ Astrophys.\  {\bf 447}, 31 (2006)
  [arXiv:astro-ph/0510447].

\bibitem{Spergel:2003cb}
  D.~N.~Spergel {\it et al.}  [WMAP Collaboration],
  Astrophys.\ J.\ Suppl.\  {\bf 148}, 175 (2003)
  [arXiv:astro-ph/0302209]. 
%
  D.~N.~Spergel {\it et al.}  [WMAP Collaboration],
  Astrophys.\ J.\ Suppl.\  {\bf 170}, 377 (2007)
  [arXiv:astro-ph/0603449].

\bibitem{Tegmark:2006az}
  M.~Tegmark {\it et al.}  [SDSS Collaboration],
  Phys.\ Rev.\  D {\bf 74}, 123507 (2006)
  [arXiv:astro-ph/0608632].

\bibitem{Dolgov:2004xu}
  A.~D.~Dolgov,
  arXiv:hep-ph/0405089.

\bibitem{reviews}
J.~Frieman, M.~Turner and D.~Huterer,
  Ann.\ Rev.\ Astron.\ Astrophys.\  {\bf 46}, 385 (2008);
  [arXiv:0803.0982 [astro-ph]];
E.~J.~Copeland, M.~Sami and S.~Tsujikawa,
  Int.\ J.\ Mod.\ Phys.\  D {\bf 15}, 1753 (2006);
 [arXiv:hep-th/0603057];
P.~J.~E.~Peebles and B.~Ratra,
  Rev.\ Mod.\ Phys.\  {\bf 75}, 559 (2003);
  [arXiv:astro-ph/0207347];
R.~R.~Caldwell and M.~Kamionkowski,
  Ann.\ Rev.\ Nucl.\ Part.\ Sci.\  {\bf 59}, 397 (2009).
  [arXiv:astro-ph/0903.0866].

\bibitem{Vissani:2009vv}
  F.~Vissani, G.~Pagliaroli and F.~L.~Villante,
  Nuovo Cim.\  {\bf 032C}, 353 (2010)
  [arXiv:astro-ph/0912.4580].

\bibitem{Bethe:1984ux}
  H.~A.~Bethe and J.~R.~Wilson,
  Astrophys.\ J.\  {\bf 295}, 14 (1985).

\bibitem{hung}
   P.~Q.~Hung,
  [arXiv:hep-ph/0010126].

\bibitem{Fardon:2003eh}
  R.~Fardon, A.~E.~Nelson and N.~Weiner,
  JCAP {\bf 0410}, 005 (2004)
  [arXiv:astro-ph/0309800].

\bibitem{Kaplan:2004dq}
  D.~B.~Kaplan, A.~E.~Nelson and N.~Weiner,
  Phys.\ Rev.\ Lett.\  {\bf 93}, 091801 (2004)
  [arXiv:hep-ph/0401099].

\bibitem{Afshordi:2005ym}
  N.~Afshordi, M.~Zaldarriaga and K.~Kohri,
  Phys.\ Rev.\  D {\bf 72}, 065024 (2005)
  [arXiv:astro-ph/0506663].

\bibitem{Bjaelde:2007ki}
  O.~E.~Bjaelde, A.~W.~Brookfield, C.~van de Bruck, S.~Hannestad, D.~F.~Mota, L.~Schrempp and D.~Tocchini-Valentini,
  JCAP {\bf 0801}, 026 (2008)
  [arXiv:astro-ph/0705.2018].

\bibitem{Barger:2005mn}
  V.~Barger, P.~Huber and D.~Marfatia,
  Phys.\ Rev.\ Lett.\  {\bf 95}, 211802 (2005)
  [arXiv:hep-ph/0502196];
  V.~Barger, D.~Marfatia and K.~Whisnant,
  Phys.\ Rev.\  D {\bf 73}, 013005 (2006)
  [arXiv:hep-ph/0509163].

\bibitem{Cirelli:2005sg}
  M.~Cirelli, M.~C.~Gonzalez-Garcia and C.~Pena-Garay,
  Nucl.\ Phys.\  B {\bf 719}, 219 (2005)
  [arXiv:hep-ph/0503028].

\bibitem{GonzalezGarcia:2005xu}
  M.~C.~Gonzalez-Garcia, P.~C.~de Holanda and R.~Zukanovich Funchal,
  Phys.\ Rev.\  D {\bf 73}, 033008 (2006)
  [arXiv:hep-ph/0511093].

\bibitem{Gu:2005eq}
  P.~H.~Gu, X.~J.~Bi and X.~m.~Zhang,
  Eur.\ Phys.\ J.\  C {\bf 50}, 655 (2007)
 [arXiv:hep-ph/0511027].

\bibitem{Gu:2005pq}
  P.~H.~Gu, X.~J.~Bi, B.~Feng, B.~L.~Young and X.~Zhang,
  Chin.\ Phys.\  C {\bf 32}, 530 (2008)
  [arXiv:hep-ph/0512076].

\bibitem{Schwetz:2005fy}
  T.~Schwetz and W.~Winter,
  Phys.\ Lett.\  B {\bf 633}, 557 (2006)
  [arXiv:hep-ph/0511177].

\bibitem{Abe:2008zza}
  K.~Abe {\it et al.}  [Super-Kamiokande Collaboration],
  Phys.\ Rev.\  D {\bf 77}, 052001 (2008)
  [arXiv:hep-ex/0801.0776].

\bibitem{Zurek:2004vd}
  K.~M.~Zurek,
  JHEP {\bf 0410}, 058 (2004)
  [arXiv:hep-ph/0405141].

\bibitem{holanda}
  P.~C.~de Holanda,
  JCAP {\bf 0907}, 024 (2009)
  [arXiv:hep-ph/0811.0567].

\bibitem{Li:2005zd}
  H.~Li, B.~Feng, J.~Q.~Xia and X.~Zhang,
  Phys.\ Rev.\  D {\bf 73}, 103503 (2006)
  [arXiv:astro-ph/0509272].

\bibitem{Komatsu:2010fb}
  E.~Komatsu {\it et al.},
  arXiv:astro-ph/1001.4538].
  Also for more information see website {\it http://map.gsfc.nasa.gov/news/index.html}

\bibitem{Kusenko:2007wv}
  A.~Kusenko,
  AIP Conf.\ Proc.\  {\bf 917}, 58 (2007)
  [arXiv:hep-ph/0703116].

\bibitem{Nunokawa:1997ct}
  H.~Nunokawa, J.~T.~Peltoniemi, A.~Rossi and J.~W.~F.~Valle,
  Phys.\ Rev.\  D {\bf 56}, 1704 (1997)
  [arXiv:hep-ph/9702372].

\bibitem{nuas}
  J.~Fetter, G.~C.~McLaughlin, A.~B.~Balantekin and G.~M.~Fuller,
  Astropart.\ Phys.\  {\bf 18}, 433 (2003)
  [arXiv:hep-ph/0205029].
  G.~C.~McLaughlin, J.~M.~Fetter, A.~B.~Balantekin and G.~M.~Fuller,
  Phys.\ Rev.\  C {\bf 59}, 2873 (1999)
  [arXiv:astro-ph/9902106].

\bibitem{Peccei:2004sz}
  R.~D.~Peccei,
  Phys.\ Rev.\  D {\bf 71}, 023527 (2005)
  [arXiv:hep-ph/0411137].

\bibitem{minakata}
  H.~Minakata, H.~Nunokawa, R.~Tomas and J.~W.~F.~Valle,
  JCAP {\bf 0812}, 006 (2008)
  [arXiv:hep-ph/0802.1489].

\bibitem{Cirelli:2004cz}
  M.~Cirelli, G.~Marandella, A.~Strumia and F.~Vissani,
  Nucl.\ Phys.\  B {\bf 708}, 215 (2005)
  [arXiv:hep-ph/0403158].

\bibitem{msw}
  L.~Wolfenstein,
  Phys.\ Rev.\  D {\bf 17}, 2369 (1978);
  S.~P.~Mikheev and A.~Y.~Smirnov,
  Sov.\ J.\ Nucl.\ Phys.\  {\bf 42}, 913 (1985)
  [Yad.\ Fiz.\  {\bf 42}, 1441 (1985)];
  S.~P.~Mikheev and A.~Yu.~Smirnov,
  Nuovo Cim.\  C {\bf 9}, 17 (1986).

\bibitem{Kainulainen:1990bn}
  K.~Kainulainen, J.~Maalampi and J.~T.~Peltoniemi,
  Nucl.\ Phys.\  B {\bf 358} (1991) 435.

\bibitem{amol}
  B.~Dasgupta and A.~Dighe,
  Phys.\ Rev.\  D {\bf 77}, 113002 (2008)
  [arXiv:hep-ph/0712.3798].

\bibitem{friedland}
  A.~Friedland,
  Phys.\ Rev.\ Lett.\  {\bf 104}, 191102 (2010)
  [arXiv:hep-ph/1001.0996].

\bibitem{duan}
  H.~Duan, G.~M.~Fuller, J.~Carlson and Y.~Z.~Qian,
  Phys.\ Rev.\ Lett.\  {\bf 97}, 241101 (2006)
  [arXiv:astro-ph/0608050].

\bibitem{dasgupta}
  B.~Dasgupta, A.~Dighe, G.~G.~Raffelt and A.~Yu.~Smirnov,
  Phys.\ Rev.\ Lett.\  {\bf 103}, 051105 (2009)
  [arXiv:hep-ph/0904.3542].

\bibitem{Giuntibook}
  C.~Giunti and C.~W.~Kim,
  {\it Fundamentals of Neutrino Physics and Astrophysics}
(Oxford University Press, Oxford, 2007).

\bibitem{Keil:2002in}
  M.~T.~Keil, G.~G.~Raffelt and H.~T.~Janka,
  Astrophys.\ J.\  {\bf 590}, 971 (2003)
  [arXiv:astro-ph/0208035].

\bibitem{totani}
  T.~Totani, K.~Sato, H.~E.~Dalhed and J.~R.~Wilson,
  Astrophys.\ J.\  {\bf 496}, 216 (1998)
  [arXiv:astro-ph/9710203].

\bibitem{garching}
  M.~Rampp and H.~T.~Janka,
  Astron.\ Astrophys.\  {\bf 396}, 361 (2002)
  [arXiv:astro-ph/0203101].

\bibitem{Franca:2009xp}
  U.~Franca, M.~Lattanzi, J.~Lesgourgues and S.~Pastor,
  Phys.\ Rev.\  D {\bf 80}, 083506 (2009)
  [arXiv:astro-ph/0908.0534].

\bibitem{k}
K.S.~Hirata {\it et al.} [Kamiokande-II Collaboration],
  Phys.\ Rev.\  D {\bf 38} (1988) 448;
K.~Hirata {\it et al.}  [Kamiokande-II Collaboration],
  Phys.\ Rev.\ Lett.\  {\bf 58} (1987) 1490.

\bibitem{i}
  R.M.~Bionta {\it et al.} [IMB Collaboration],
  Phys.\ Rev.\ Lett.\  {\bf 58} (1987) 1494;
C.B.~Bratton {\it et al.}  [IMB Collaboration],
  Phys.\ Rev.\  D {\bf 37} (1988) 3361.

\bibitem{b}
E.N.~Alekseev, L.N.~Alekseeva, V.I.~Volchenko and I.V.~Krivosheina [Baksan Collaboration],
  JETP Lett.\  {\bf 45} (1987) 589;
  [Pisma Zh.\ Eksp.\ Teor.\ Fiz.\  {\bf 45}, 461 (1987)].
E.N. Alekseev, L.N.~Alekseeva, I.V.~Krivosheina and V.I.~Volchenko [Baksan Collaboration],
  Phys.\ Lett.\  B {\bf 205} (1988) 209.

\bibitem{murayama}
  H.~Murayama and T.~Yanagida,
  Phys.\ Lett.\  B {\bf 520}, 263 (2001)
  [arXiv:hep-ph/0010178].

\bibitem{pagliaroli}
  G.~Pagliaroli, F.~Vissani, M.~L.~Costantini and A.~Ianni,
  Astropart.\ Phys.\  {\bf 31}, 163 (2009)
  [arXiv:astro-ph/0810.0466].

\bibitem{qian}
  Y.~Z.~Qian, G.~M.~Fuller, G.~J.~Mathews, R.~W.~Mayle, J.~R.~Wilson and S.~E.~Woosley,
  Phys.\ Rev.\ Lett.\  {\bf 71}, 1965 (1993).

\bibitem{rprocessreview}
  M.~Arnould, S.~Goriely and K.~Takahashi,
  Phys.\ Rept.\  {\bf 450}, 97 (2007)
  [arXiv:astro-ph/0705.451].

\bibitem{RossiTorres:2010zz}
  F.~Rossi-Torres, M.~M.~Guzzo and P.~C.~de Holanda,
  J.\ Phys.\ Conf.\ Ser.\  {\bf 203}, 012141 (2010).

\end{thebibliography}
\end{document}